\documentclass[a4paper,11pt]{article}
\usepackage{graphicx}
\usepackage{amsfonts}
\usepackage{amsmath}
\usepackage{hyperref}

\begin{document}

\title{Who Needs Trust for 5G?}

\author{Chris J Mitchell,\\
Royal Holloway, University of London, Egham TW20 0EX, UK\\
\url{www.chrismitchell.net}
}

\maketitle

\begin{abstract}
There has been much recent discussion of the criticality of the 5G infrastructure, and whether
certain vendors should be able to supply 5G equipment. The key issue appears to be about trust,
namely to what degree the security and reliability properties of 5G equipment and systems need to
be trusted, and by whom, and how the necessary level of trust might be obtained. In this paper, by
considering existing examples such as the Internet, the possible need for trust is examined in a
systematic way, and possible routes to gaining trust are described. The issues that arise when a
security and/or reliability failure actually occurs are also discussed.  The paper concludes with a
discussion of possible future ways of enabling all parties to gain the assurances they need in a
cost-effective and harmonised way.
\end{abstract}

\section{Introduction}

In recent months there has been much public discussion of the security properties of 5G mobile
telecommunications systems, and in particular whether certain suppliers should be permitted to
provide 5G equipment --- see, for example, \cite{Burgess20,Laudrain20,Sweeney20}. These, often
rather fevered, discussions have often taken place within a apparently political context. It is to
be hoped that in the longer run we can examine all the issues relating to the supply and operation
of communications equipment in a rather more dispassionate and scientific way, and it is the
author's hope that this paper will contribute in a small way to such a discussion. There are
certainly important issues to be considered, and the way forward is not always clear.

Discussions have often revolved around the word \emph{trust}, a difficult concept which forms the
main theme of this paper. Much of the recent reporting of the 5G issue has focussed on whether
suppliers can be trusted (see, for example, \cite{Lecher19,Mathieson19}).  Of course, the notion of
trust has rather emotive connotations, and so we need to try to understand better what it might
mean for one party to trust another in the context of mobile telecommunications.  This in turn
leads us to the need to consider who might be the principal actors in the global digital ecosystem,
and what are their needs for trust?  Any such analysis needs to be performed in the context of the
history of global digital communications and existing trust-building measures.

The remainder of this paper is organised as follows.  We start in
\S\ref{section-definition} by briefly examining the meaning of the word trust.
In \S\ref{section-who} we consider where trust is needed, i.e.\ in what
situations trust is required and by and in whom, and this leads naturally to
\S\ref{section-type}, where the types of trust that might be needed are
discussed.  Possible means of enabling trust are described in
\S\ref{section-enable}, and the relevance of these for 5G is addressed in
\S\ref{section-5G}.  The issue of what happens when trust fails is the subject
of \S\ref{section-wrong}. The final main section, \S\ref{section-frameworks},
contains a discussion of how possible frameworks for performing 5G trust
evaluations can support cost-effective decision-making in the future.  The
paper concludes with a brief summary in \S\ref{section-summary}.

\section{What is trust?}  \label{section-definition}

The word `trust' is very heavily used, and has a wide range of meanings --- see, for example, Voas
\cite{Voas20}. Of course, for the purposes of this paper the scope is much reduced.  We adopt here
the following definition, taken from the ISO Online Browsing
Platform\footnote{\url{https://www.iso.org/obp/ui}}
\begin{quote}
degree to which a user or other stakeholder has confidence that a product or system will behave as
intended
\end{quote}
(see ISO/IEC 25010:2011 \cite{ISO25010:2011}).  That is, trust here is about the behaviour of a
product or system.  Note that we are interested only in this rather narrow interpretation of trust
in a primarily business context --- of course, there has been a huge volume of work looking at
trust from a psychological and sociological perspective, including the work of Simmel
\cite{Simmel50}.

In the context of security, trust is about behaviour relating to the \emph{confidentiality} and
\emph{integrity} of information assets handled by the product or system, and/or the
\emph{availability} of the service provided.  It seems clear that these are precisely the issues
that have caused so much recent concern, and we therefore use this definition in the remainder of
the paper.

Even when we restrict our attention to 5G, there are many aspects to trust as defined above. The
main focus of this paper is on the aspects of trust that affect `long-term' decision-making, e.g.\
\begin{itemize}
\item what systems and products should network operators and other service providers choose to
    use to provide their services;
\item which network operators and other service providers should be used to provide important
    services, e.g. to support manufacturing or transport services.
\end{itemize}
In many cases, the appropriate level of trust can be achieved using business-related methods such
as product certification and the establishment of service level agreements and contracts (as
discussed in greater detail in \S\ref{section-enable}).

However, this ignores some very important trust-related issues, namely those affecting more
`short-term' decision-making.  Examples might include deciding whether an end user should:
\begin{itemize}
\item employ a particular service at a specific time for a given purpose, e.g.\ to send a
    security-critical message or to act on (trust) a received email;
\item trust a transaction completed using a service provided by 5G technology.
\end{itemize}
Providing the necessary level of trust for such decisions would appear to be a technology-based
issue, rather than a business/commercial one.  That is, the party required to make the decision may
wish to see evidence that the technology used to provide the service is suitable for the purpose.
Such technology might involve a wide range of techniques, including methods for achieving
distributed trust (e.g.\ block chain). A discussion of some of the key issues has been provided in
a recent survey by Ahmad et al.\ \cite{Ahmad19}. However, we do not propose to discuss such issues
further here.

\section{Who needs to trust whom?}  \label{section-who}

We start by considering the key stakeholders in any global communications system, such as the
Internet.  These stakeholders can be divided into four main categories:
\begin{itemize}
\item \emph{end users}, both organisational and individual;
\item \emph{equipment/system manufacturers, including software suppliers}, a broad category
    covering terminal manufacturers, manufacturers of network infrastructure and providers of
    software used to enable network services;
\item \emph{network service providers}, including not only `traditional' network operators but
    also a range of other network service providers; and
\item \emph{regulatory and standardisation organisations}, including governments, government
    agencies and supranational organisations.
\end{itemize}

The first three classes form the operational system, whereas the regulators monitor and exert some
control over the operation of the system from outside.  In principle at least, the trust
relationships between the operational stakeholders form a chain, with users needing to trust the
service providers, who in turn must trust both other service providers and the equipment
manufacturers.  The regulators and standardisation bodies then trust all the other parties to
perform their roles in accordance with the established rules and regulations.

\section{What type of trust is needed?}  \label{section-type}

We next consider how trust is established and managed between the various stakeholders. Perhaps
most significantly, it is important to realise that the type and degree of trust required will vary
widely depending on the nature of the network or service, and what it is being used for.  There are
two types of trust of particular relevance in a communications context.

\begin{itemize}
\item \emph{Security}:  how certain can I be that data sent and received is going to/coming
    from the right entity and has not been interfered with?
\item \emph{Reliability}: how reliable is the network and/or service, i.e. how likely is it to
    be available when needed, and what bandwidth and other service level guarantees are there?
\end{itemize}

In the case of the Internet, for end users the level of trust in security is not very high at all;
indeed, encryption at the application layer encryption or immediately below (e.g.\ as provided by
Transport Layer Security (TLS) \cite{Rescorla01}) is routinely used to protect web traffic. This is
very reasonable since end users are offered no guarantees about how their data will be routed and
what third parties might have access to it.  For example, many of us routinely use
third-party-provided Wi-Fi, e.g. in airports or coffee shops, and we have no control over (or
knowledge about) who is providing the service and who may be monitoring or interfering with the
traffic.  The only option is to use TLS and/or Virtual Private Networks (VPNs) for all sensitive
traffic. Similar issues arise with use of mobile communications networks, where trusting the
network requires not only trusting the network provider (often a completely unknown quantity when
roaming) and also trusting that the physical network infrastructure has not been compromised.

The level of \emph{reliability} for end users is also highly variable, typically depending on the
type of network used to access the Internet. For example, home users may experience much reduced
bandwidth at peak times, where a single network link is shared amongst many premises. Similarly,
the quality of access via Wi-Fi may depend on a range of factors and may vary dynamically.  For
mobile networks of all generations (including 5G), there are inherent limits to the quality of
service, since no network guarantees universal coverage and connectivity may also be lost in times
of peak demand.

As far as \emph{service providers} are concerned, the level of trust in Internet security when
working with other service providers is also likely to be very low, since the Internet relies on a
loose collaboration between entities and is not subject to tight regulatory control. The level of
security trust in equipment manufacturers is not an issue that has been addressed to any great
extent. There have been occasions when manufacturers have been accused of deliberately installing
security backdoors in their equipment (see, for example, Lee \cite{Lee12}), but this is not an
issue that has been widely explored.

Finally, the role of \emph{regulators} on the operation of the Internet appears to be very limited.
That is, the required level of trust by regulators in service providers and manufacturers is
generally low.  Of course, the Internet Engineering Task Force
(IETF)\footnote{\url{https://www.ietf.org/}} plays a key role in setting technical standards for
the Internet, and IANA\footnote{\url{https://www.iana.org/}} manages functions such as IP address
allocation. However, in general there is no regulation governing the procurement of equipment to
provide the Internet infrastructure.

In summary, at least as far as the Internet is concerned, we appear to be starting from a
relatively low trust base, i.e. users and service providers employ and operate the Internet without
high levels of trust. This situation is perhaps rather surprising when one considers the degree on
which so many aspects of modern life rely on the Internet. Instead of demanding a perhaps
unrealisable level of trustworthiness in the Internet, we rely on a range of mitigations, including
providing security on a case by case basis (e.g. using TLS or Secure Shell (SSH) \cite{RFC4251}),
and exploiting redundancy of various types --- including having a multitude of access methods, e.g.
via fixed, public Wi-Fi or mobile networks, and multiple possible routes between end points.

Whether this situation is reasonable in the long term remains to be seen.  Certainly the Internet
as currently implemented is easily made unavailable by state actors, e.g. in times of war, as past
experience shows.

\section{How do we enable trust?}  \label{section-enable}

There would appear to be three main instruments by which the necessary levels of trust can be
established:

\begin{itemize}
\item through \emph{contractual agreements and SLAs};
\item through assessments of the \emph{reputation} (trustworthiness)
    of the suppler; and
\item via \emph{product assurance} mechanisms (including product certifications).
\end{itemize}

For \emph{end users} the first two mechanisms are key as far as network service
    is concerned --- although the assurance mechanism is, of course, relevant to the provision of
    user terminals (handsets).  The level of security and reliability of service guaranteed via
    contract will typically be low.  That is, service contracts will typically not guarantee
    any particular level of availability or data throughput.  Of course, even if guarantees
are available with respect to local access, this does not mean that there are any guarantees about
end-to-end network security or reliability since this will almost certainly be out of the control
of the provider of network access.  Similar limitations apply to the use of reputation ---
certainly the usefulness of reputation as a measure of service reliability or security is very
limited; indeed, with respect to security, it may be of no value at all, since most users will have
no idea of how secure their network use is, and to be useful a service's reputation requires
existing users to be able to give an informed assessment.

For \emph{service providers}, the degree to which contractual agreements and
    reputation help them in gaining trust with respect to equipment manufacturers is moot.  One might
    reasonably assume that the terms of purchase are pretty much the same for all suppliers,
    and it is not clear to what degree the major infrastructure equipment manufacturers have
    varying reputations.  This leads naturally to consideration of product certification.

Product assurance via certification has many different aspects.  Certain products and services are
critical for wider society and, as a result, the need for certification is very high; however, the
failure of other products might be unfortunate, but less serious, and hence the need for
certification is lower. The degree to which a purchaser depends on the technology should be
established before the need for product certification can be assessed.

There is clearly a need for purchasers of telecommunications products to be assured regarding their
security properties, where products include handsets, USIMs (including embedded and virtual) and
infrastructure.  This typically involves certifications of products against standards, a role the
Global Certification Forum (GCF)\footnote{\url{https://www.globalcertificationforum.org/}} has
performed for parts of the mobile industry.  However, GCF certifications are mainly focussed on
testing functionality/interoperability, rather than providing guarantees about security resilience.

More generally, over the last 40 years standardised techniques and processes have been developed to
enable consumers to gain confidence in the security properties of IT products and systems --- for a
helpful review see Chapter 16 of Kizza \cite{Kizza17}. Of particular importance are the Common
Criteria standards (ISO/IEC 15408 \cite{ISO15408-1:2009,ISO15408-2:2008,ISO15408-3:2008}) which
specify how testing laboratories can test and certify products so that purchasers and users can be
confident that: (a) products do what they should, and, perhaps even more importantly, (b) they
don't do what they should not.  However, gaining certification for a product under the common
criteria can be costly, time-consuming, and there are problems with mutual recognition of
certifications --- see, for example, Kallberg \cite{Kallberg12}.  As a result, new
mobile-industry-specific approaches are being developed by 3GPP
(SECAM/SCAS)\footnote{\url{https://www.3gpp.org/news-events/3gpp-news/1569-secam_for_3gpp_nodes}}
and GSMA
(NESAS)\footnote{\url{https://www.gsma.com/security/network-equipment-security-assurance-scheme/}}
to provide evaluation processes suitable for 5G products and systems.

Recent public concern over the security properties of certain 5G systems emphasises the lack of a
universally agreed method by which assurance can be gained in the security properties of products.
National and vendor-specific approaches, such as the UK's industry-leading Huawei Cyber Security
Evaluation Centre (HCSEC) (see the 2019 annual report \cite{HCSEC19}), are clearly not globally
scalable, not least because of the stringent requirements, high cost and major delays inherent in
such an approach.  As a result internationally recognised, generally applicable, approaches are
needed.

In the context of 5G, the hardware is often discussed, but it is actually the software that is the
real key component of the network. This constitutes a problem when certifying a product, as
software changes regularly, not least as a result of automatic updates and patches.  Certification
will never be the silver bullet that will address the trust issue once and for all, but could be of
great help providing it is conducted appropriately.

Finally, it is also worth mentioning the notion of \emph{zero trust networks}, as discussed in
detail in draft NISP SP 800-207 \cite{NIST800-207_20}.  As stated there, `zero trust security
models assume that an attacker is present in the network' so that security measures include
`minimizing access to resources (such as data and compute resources and applications) to only those
users and assets identified as needing access as well as continually authenticating and authorizing
the identity and security posture of each access request'.  Clearly adoption of a zero trust
approach can reduce the need to trust individual elements of a large network whilst increasing
resilience, but (a) the architecture needs to be built into the specifications for system
interfaces, and (b) the need for additional authentication and authorisation will almost inevitably
increase complexity and cost.  This appears to be very much a long-term solution, that has
relatively little relevance for 5G\@.  Indeed, the ideas are still very much at an
experimental/research stage --- see, for example, Eidle et al.\ \cite{Eidle17}.

\section{Trust for 5G}  \label{section-5G}

It is clear that, until now, the level of trust in the security and reliability of the Internet,
and in particular in mobile telecommunications, is low.  Whilst this may sound disturbing, the good
news is that this has not prevented many of us enjoying the huge range of applications built on the
Internet in many aspects of our daily lives.  We have obtained the degree of trust we need by
building mitigations on top of the base level of communications that are provided.  Of course, as
we are told very regularly, 5G changes everything, and so we need to consider how our trust
requirements will evolve as 5G becomes more pervasive. Some of the key trust issues for 5G are
summarised in the recent article by Kshetri and Voas \cite{Kshetri20}.

5G promises to become fundamental to many functions of society, including automated manufacturing,
autonomous vehicles, and edge computing.  More specifically, a much wider range of industry
verticals, each with different value chains, will be involved in 5G than in past generations of
mobile telecommunications. Moreover, the historic concentration on business to consumer (B2C)
applications will expand to include many different types of business to business (B2B)
applications, including deployment in critical infrastructure operations. In the fields of car
sharing and mobility provision, for example, the 5G-connected ecosystem will span providers of
rental cars and mobility services, manufacturers of cars and mobile terminals, mobile
telecommunications operators, application providers, and the manufacturers of smart cards. These
changes will create major new challenges for governments, industry organisations, individual
ecosystem members (operators, vendors, etc.), and academia.

For governments, the most significant question is whether and how they should regulate the 5G
industry. What should be decided by government and what should be decided by the market? For
industry organizations, many questions arise regarding the boundaries of responsibilities,
expectations, power and rights for each stakeholder in the ecosystem. As a result, the degree to
which society relies on its existence will continue to increase.

The increased need for assurance in system and product functionality has given rise to a major
growth of interest in, and development of, methods of certification of telecommunications products
and systems.  One reason why certification has recently arisen as a major issue stems from the
belief that 5G is transformative, and that citizens in Europe and around the world will be
increasingly reliant on this technology.  As a result, the need for trust and certification is
becoming progressively higher.  Whilst this is certainly not a negative development, it is a little
curious that the question of certification has not been a major issue previously.  Indeed, 4G is
also widely relied on, and yet there is no unified European certification scheme for it.

As far as recent developments on certification are concerned, the EU Cybersecurity Act
\cite{CybersecurityAct19} provides a framework for the European Union to harmonise the way
certification is conducted; it will probably also be adopted by those outside the EU who wish to
follow a similar path.  For 5G specifically, there are a number of initiatives, including GSMA's
NESAS and the work of 3GPP including in particular the creation of a range of SCAS's (Security
Assurance Specifications for specific products).  More debate is needed on how the different
aspects can fit together, what the levels of criticalities are, and the degree to which
certification is needed. Ideally  a harmonised base-line of certification, based on the
Cybersecurity Act for both 5G infrastructure and consumer products, would  be established across
the EU. This could then be built on at the national level, as well as providing a reference level
for countries outside the EU\@. Of course, there is still a long way to go to reach this aim,
although many of us hope that a base-line level of certification will be in place sooner than one
might think.

In parallel to the use of certification as a way of enhancing trust, another key approach involves
diversification of supply.  That is, if equipment can be procured from more than supplier, then the
odds of the entire network failing simultaneously is much reduced.  Of course, the degree to which
this can reduce the risk of a major failure depends on the nature of the equipment; some functions
are critical to the entire network, and if they fail then the entire network fails.  Generally
speaking, trying to minimise the risk of reliability failure requires redundancy, which in turn
increases cost.  This raises an interesting question --- if regulators require service providers to
install more equipment than necessary to provide a service, and/or purchase equipment from multiple
vendors, then this will clearly increase costs for the service provider; who will pay?

Some regulatory bodies have already taken the decision to require 5G service providers to avoid
over-dependency on a single supplier, notably in the case of the UK government's recent decision to
limit one supplier to 35\% of the total in a network (see Sweeney, \cite{Sweeney20}).  If course,
this decision could also be seen as a move to provide financial support to other suppliers.

Finally, it is interesting to observe that the drive for certification schemes as a means of
enhancing trust in the provision of network equipment appears to be coming from regulatory and
other government bodies, rather than from service providers --- i.e.\ the actual purchasers of
equipment.  Many government bodies see their role as to require service providers to ensure a high
level of trustworthiness in their service, above what they might need to meet their contractual
obligations to customers.  This presumably arises from a belief that that the mobile communications
infrastructure is a critical national infrastructure, and hence something whose reliability must be
protected by government.  If this assumption is true, it is surprising that similar considerations
are not applied to the other parts of the Internet communications system, since if any significant
part of the Internet fails then the effects will be very significant.

\section{What happens when things go wrong?}  \label{section-wrong}

We conclude this discussion of trust by considering what happens when things go wrong; that is,
what happens when the expected level of security or reliability is not met.  Who is responsible for
picking up the pieces?  Of course, this depends on which party has been let down, although we focus
here primarily on the service providers, since end users have little recourse apart from pursuing
claims for compensation because of a failure to meet contractual obligations.

In general, this is a key issue relating to trust.  One can think of trust as influencing decisions
as to what equipment and/or software to buy (notably by service providers), and also influencing
decisions by regulators as to what equipment service providers and others are permitted to buy.
However, once the systems have been procured and are operating, we also need to think about what
might happen if the equipment fails, where failure here covers both ceasing to function (loss of
reliability) and loss of security, e.g.\ enabling large scale leaking of data.

Whilst a reliability breach will be immediately obvious, a security breach may be much less so.
Nonetheless, if data is leaked by equipment on a large scale then this is likely to be detected. Of
course, such matters are typically managed through contracts, and in the event of the failure of a
single piece of equipment this is clearly something that can be managed in the usual way. However,
a major concern for all parties must be if a large number of pieces of equipment all fail at the
same time.  This could happen in many ways, e.g.\ as a result of criminal hackers or state actors
exploiting vulnerabilities in software.

There would thus be two main ways in which an equipment manufacturer might ultimately be
responsible for such a breach:
\begin{itemize}
\item through negligence, e.g.\ by leaving an exploitable vulnerability in a product because of
    poor software engineering practices, or
\item through deliberately including some kind of back door in equipment which enables a third
    party to gain unauthorised access to data or to cause equipment to stop functioning.
\end{itemize}
After a major breach investigations may reveal which of the two is the cause (although it is
possible that a back door could be engineered to look like an accidental vulnerability).

If the breach is very serious, the reputation of the supplier is likely to be badly damaged
whichever of the two causes is true.  Of course, if an equipment manufacturer is ever found to be
guilty of deliberately enabling a security or reliability failure, then their reputation and hence
their trust will be damaged possibly beyond repair.  However, if the system that fails is `critical
infrastructure' then maybe the damage will already have been done, which explains the concerns of
some governments and regulators.  Their primary concern is probably to ensure that the possibility
and scope of major failures can be minimised, \emph{in advance} of any problems.

\section{Possible trust frameworks}  \label{section-frameworks}

There are well-established risk-based ways of determining how to achieve appropriate trade-offs
between the need for security and the desire to make cost-effective investments --- see, for
example, the ISO/IEC 27001 information security management system \cite{ISO27001:2013}. The
fundamental idea is to base all decisions on an assessment of risk.  In the context of trust, this
means deciding on the degree of trust required in a 5G system component based on an assessment of
the risks arising from a security failure.

For example, a system component that is critical to the operation of an entire network would have a
high level of risk associated with it, and would therefore require a high level of trust in the
component.  As a result, a manufacturer might be required to obtain certification for the product
or system which involved rigorous examination of the hardware and software, and the reputation of
the supplier would also need to be high. Alternatively, a product with minimal access to sensitive
data (meaning that confidentiality and integrity risks are low), and with minimal impact on the
availability of the system, might only require a low level of trust.  In principle, such
evaluations could be performed by every acquirer of equipment, but in practice it is unreasonable
to expect every party to have the necessary expertise.

How can such a situation be addressed?  One possibility would be to define a
standardised framework for making such assessments. This could have a number of
advantages.

\begin{itemize}
\item It would de-skill the task of evaluating the level of trust required
    in a system to be procured.
\item It would mean that the process of trust evaluation could be done in a
    uniform way across the industry, meaning that regulators could rely on
    individual companies making decisions in a recognised way.
\item Regulators could stipulate requirements for product assurance to all
    organisations within its remit, using a standardised terminology and
    framework.
\item It would help to support standardisation of the means of acquiring
    trust, e.g.\ through product or system evaluations and certifications,
    or through provision of evidence regarding the product development
    lifecycle.  Such standardisation would potentially significantly reduce
    costs for equipment and system manufacturers, in that they would not be
    required to provide evidence to enable trust in a different form for
    each customer.
\end{itemize}

One such framework is currently at an early stage of development, namely ITU-T
X.5GSec-t\footnote{https://www.itu.int/md/T17-SG17-200317-C/en}, \emph{Security framework based on
trust relationship for 5G ecosystem}, being produced by SG17.  However, unfortunately, the latest
draft from March 2020\footnote{https://www.itu.int/md/T17-SG17-C-0821/en} is not publicly
available.  The role of this draft standard, work on which started in 2018, is discussed in a 2019
presentation \cite{Youm19}.

There are potentially three different facets of such a framework.
\begin{itemize}
\item \emph{Governmental Responsibilities}: What should be managed, controlled, planned by
    governments, and what should be decided by the market?  Are there any fundamental
    principles that each government needs to follow?
\item \emph{Industry Organisation Responsibilities}: To what degree and how should industry
    organisations work within an industry ecosystem governance framework? How should the
    framework manage the boundaries of responsibility, power and rights of each stakeholder in
    the industry ecosystem?
\item \emph{Technology}: If governments and industry organisations are able to find ways to
    work together effectively, perhaps the technology-specific aspects will prove comparatively
    easy (or at least not too difficult)? Within a system with clear boundaries of
    responsibility, power and rights, maybe the market can choose the most appropriate security
    solutions based on customer requirements?
\end{itemize}
There are clearly more questions here than answers, and much further work is needed to resolve some
of these key issues.

\section{Summary and conclusions}  \label{section-summary}

We have considered the role of trust in relationships between key parties in the 5G ecosystem; we
have also examined ways in which such trust can be provided.  In general it is likely to be
difficult for the large numbers of players in the ecosystem to understand the degree to which they
need to trust their suppliers, and also how they might gain the degree of trust that they require.
The absence of the appropriate levels of knowledge and trust could seriously distort the market,
and costs for all parties could be significantly higher than they need to be.  In conclusion, the
notion  of a framework for developing an understanding the necessary trust levels, and the way in
which trust can be developed, was introduced, and the existence of efforts in the standardisation
community to meet this need was noted.  This is clearly an area where considerably more work is
urgently needed.


\end{document}